\documentclass[11pt]{article}

\usepackage{fullpage,epsfig,amssymb,amsmath,amsfonts,amssymb,
graphicx,amsthm,algorithm,latexsym,picinpar,epstopdf}

\newcommand{\mycase}[1]{\mbox{{\underline{Case #1}}:\/}}

\newcommand{\half}{{\mbox{$\frac{1}{2}$}}}
\newcommand{\threehalfs}{{\mbox{$\frac{3}{2}$}}}

\newcommand{\onethird}{{\mbox{$\frac{1}{3}$}}}

\newcommand{\onefourth}{{\mbox{$\frac{1}{4}$}}}

\newcommand{\braced}[1]{{ \left\{ #1 \right\} }}

\newcommand{\floor}[1]{{ \lfloor #1 \rfloor }}
\newcommand{\ceiling}[1]{{ \lceil #1 \rceil }}

\newcommand{\optgain}{{\mbox{\it gain}^\ast}}

\newcommand{\gain}{\mbox{\it gain}}

\newcommand{\assign}{{\,\leftarrow\,}}

\newcommand{\NP}{{\mbox{\sf NP}}}
\newcommand{\MAXSNP}{{\mbox{\sf MAX-SNP}}}

\newcommand{\Greedy}{{\mbox{\sc Greedy}}}
\newcommand{\OneGreedy}{{\mbox{\sc 1-Greedy}}}
\newcommand{\TwoGreedy}{{\mbox{\sc 2-Greedy}}}
\newcommand{\Clique}{{\mbox{\sc Clique}}}
\newcommand{\FlowMntrs}{{\mbox{\sc FlowMntrs}}}
\newcommand{\WFlowMntrs}{{\mbox{\sc WFlowMntrs}}}
\newcommand{\DecFlowMntrs}{{\mbox{\sc DecFlowMntrs}}}

\newcommand{\bridges}{{\mbox{\it Br}}}

\newcommand{\resG}{{G^\ast}}
\newcommand{\resV}{{V^\ast}}
\newcommand{\resE}{{E^\ast}}

\newtheorem{theorem}{Theorem}

\newtheorem{lemma}[theorem]{Lemma}

\newtheorem{observation}[theorem]{Observation}

\newcommand{\ignore}[1]{}

\begin{document}

%%%%%%%%%%%%%%%%%%%%%%%%%%%%%%%%%%%%%%%%%%%%%%%%%%%%%%%%%%%%%%%%%%%%%%
%%%%%%%%%%%%%%%%%%%%%%%%%%%%%%%%%%%%%%%%%%%%%%%%%%%%%%%%%%%%%%%%%%%%%%
%%%%%%%%%%%%%%%%%%%%%%%%%%%%%%%%%%%%%%%%%%%%%%%%%%%%%%%%%%%%%%%%%%%%%%

\title{Algorithms for Placing Monitors\\ in a Flow Network}

\author{
	Francis Chin\thanks{%
		Department of Computer Science,
		The University of Hong Kong,
		Pokfulam, Hong Kong.
	  Research supported in parts by grant (HKU 7113/07E).
		}
	\and
       Marek Chrobak\thanks{%
       Department of Computer Science,
       University of California,
       Riverside, CA 92521.
       Research supported by NSF Grant CCF-0729071.
       }
       \and
	 Li Yan\footnotemark[2]
   }

\maketitle

\begin{abstract}
  In the Flow Edge-Monitor Problem, we are given an
  undirected graph $G=(V,E)$, an integer $k > 0$ and some
  unknown circulation $\psi$ on $G$. We want to find a set
  of $k$ edges in $G$, so that if we place $k$ monitors on
  those edges to measure the flow along them, the total
  number of edges for which the flow can be uniquely
  determined is maximized. In this paper, we first show that
  the Flow Edge-Monitor Problem is {\NP}-hard, and then we
  give two approximation algorithms: a $3$-approximation
  algorithm with running time $O((m+n)^2)$ and a
  $2$-approximation algorithm with running time $O((m+n)^3)$,
  where $n = |V|$ and $m=|E|$.
\end{abstract}

%%%%%%%%%%%%%%%%%%%%%%%%%%%%%%%%%%%%%%%%%%%%%%%%%%%%%%%%%%%%%%%%%%%%%%
%%%%%%%%%%%%%%%%%%%%%%%%%%%%%%%%%%%%%%%%%%%%%%%%%%%%%%%%%%%%%%%%%%%%%%
%%%%%%%%%%%%%%%%%%%%%%%%%%%%%%%%%%%%%%%%%%%%%%%%%%%%%%%%%%%%%%%%%%%%%%

\section{Introduction}
\label{sec: introduction}

We study the \emph{Flow Edge-Monitor Problem} (\emph{\FlowMntrs}, for short),
where the objective is to find $k$ edges in an undirected
graph $G=(V,E)$ with an unknown circulation $\psi$, so that if we 
place $k$
flow monitors on these edges to measure the flow along them,
we will maximize the total number of edges for which the value
and direction of $\psi$
is uniquely determined by the flow conservation
property.  Intuitively, the objective is to maximize the
number of bridge edges in the subgraph induced by edges not
covered by monitors. (For a more rigorous definition of the
problem, see Section~\ref{sec: preliminaries}.)

\begin{figure}[ht]
\hfill
    \begin{minipage}[c]{0.65\linewidth}
	\includegraphics[width=\linewidth]{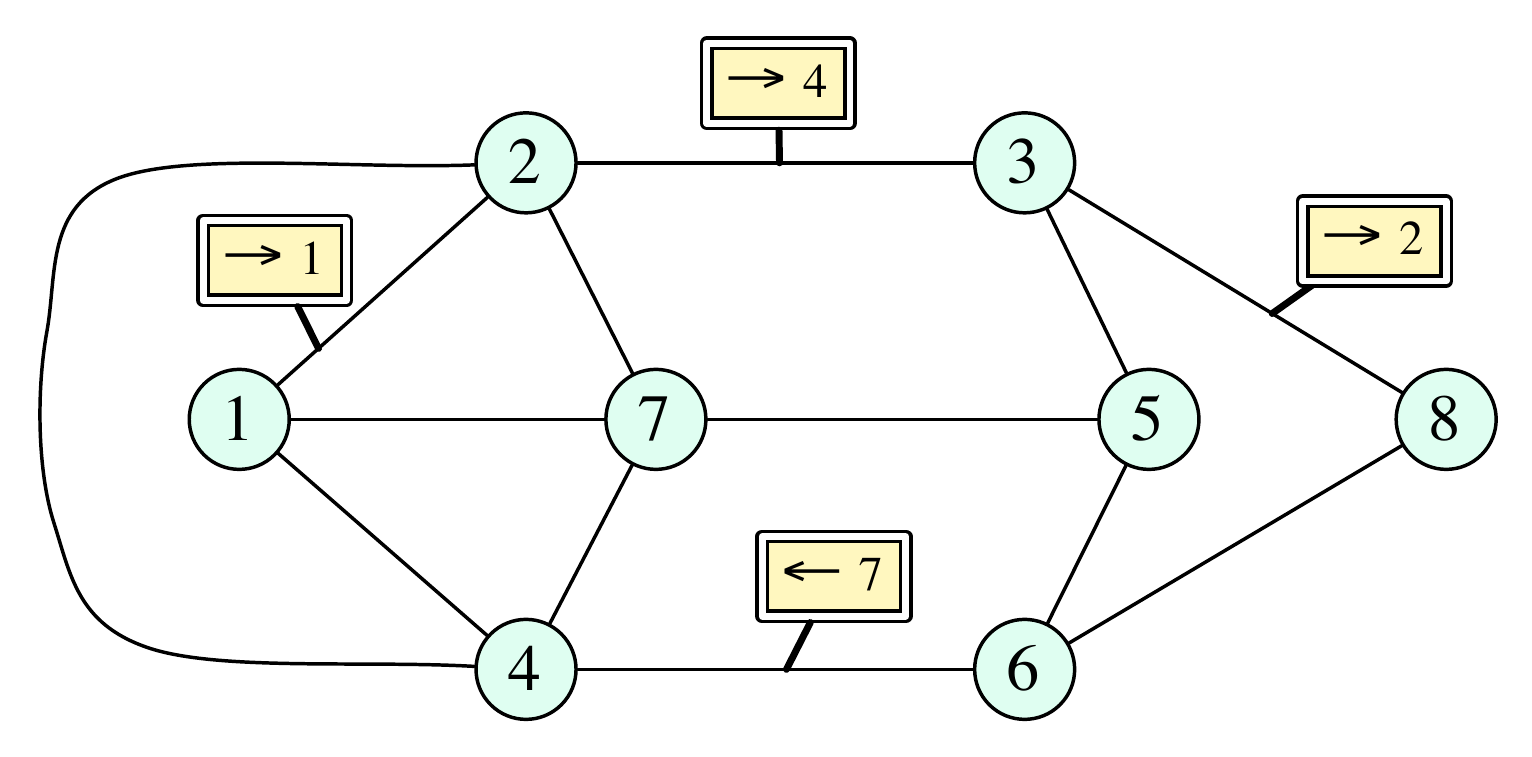}
    \end{minipage}
\hskip 0.15in
    \begin{minipage}[c]{0.25\linewidth}
    	\caption{A graph with $4$ monitors.}
		\label{fig: graph monitors example}
    \end{minipage}
\hfill{\ }
\end{figure}

Consider, for example, the graph and the monitors shown in
Figure~\ref{fig: graph monitors example}.
In this example we have $k=4$ monitors represented by
rectangles attached to edges, with measured flow values and
directions shown inside.  Thus we have $\psi(2,3) = 4$,
$\psi(3,8) = 2$, $\psi(6,4) = 7$ and $\psi(1,2) = 1$.  From
the flow conservation property, we can then determine that
$\psi(3,5) = 2$, $\psi(8,6) = 2$, $\psi(7,5) = 3$ and
$\psi(5,6) = 5$.  Thus with $4$ monitors we can determine
flow values on $8$ edges.

%%%%%%%%%%%%%%%%%%%%%%%%%%%%

\paragraph{Our results.}  We first show that the {\FlowMntrs} problem
is {\NP}-hard.  Next, we study
polynomial-time approximation algorithms. We introduce an
algorithm called {$\sigma$-\Greedy} that, in each step,
places up to $\sigma$ monitors in such a way that the number
of edges with known flow is maximized.  We then prove that
{\OneGreedy} is a $3$-approximation algorithm and that
{\TwoGreedy} is a $2$-approximation algorithm.  The running
times of these two algorithms are $O((m+n)^2)$ and $O((m+n)^3)$,
respectively, where $n = |V|$ and $m=|E|$. In both cases,
our analysis is tight. In fact, our approximation results
are stronger, as they apply to the weighted case, where the
input graph has weights on edges, and the objective is to
maximize the total weight of the edges with known flow.

%%%%%%%%%%%%%%%%%%%%%%%%%%%

\paragraph{Related work.}  A closely related problem
was studied by Gu and Jia \cite{GJ-traffic-05} who
considered a traffic flow network with directed edges. They
observed that $m-n+1$ monitors are necessary to determine the 
flow on all edges of a strongly connected graph,
and that this bound can be
achieved by placing flow monitors on edges in the complement
of a spanning tree. (The same bound applies to connected
undirected graphs.) Khuller {et al.}~\cite{KBP-meters-03}
studied an optimization problem where
pressure meters may be placed on nodes of a flow network.
An edge whose both endpoints have a
pressure meter will have the flow determined using the
pressure difference, and other edges may have the flow
determined via flow conservation property. The goal is to
compute the minimum number of meters needed to determine the
flow on every edge in the network. They showed that this
problem is {\NP}-hard and {\MAXSNP}-hard, and that a
local-search based algorithm achieves $2$-approximation. For
planar graphs, they have a polynomial-time approximation
scheme.  The model in \cite{KBP-meters-03} differs from ours
in that it assumes that the flow satisfies Kirchhoff's
current and voltage laws, while we only assume the current
law (that is, the flow preservation property).  This
distinction is reflected in different choices of ``meters":
vertex meters in \cite{KBP-meters-03} and edge monitors in our paper. 
Recall that, as explained above, minimizing the number of 
\emph{edge monitors}
needed to determine the flow on all edges is trivial, providing 
a further justification for our choice of the objective function.

The {\FlowMntrs} problem is also related to the
classical $k$-cut and multi-way cut problems
 \cite{SV-cuts-91,vazirani,DJPSY-mcuts-92}, where
the goal is to find a minimum-weight set
of edges that partitions the graph into $k$ connected
components. One way to view our monitor problem is that we want to
maximize the number of connected components obtained from
removing the monitor edges and the resulting bridge edges.

%%%%%%%%%%%%%%%%%%%%%%%%%%%%%%%%%%%%%%%%%%%%%%%%%%%%%%%%%%%%%%%%%%%%%%
%%%%%%%%%%%%%%%%%%%%%%%%%%%%%%%%%%%%%%%%%%%%%%%%%%%%%%%%%%%%%%%%%%%%%%
%%%%%%%%%%%%%%%%%%%%%%%%%%%%%%%%%%%%%%%%%%%%%%%%%%%%%%%%%%%%%%%%%%%%%%

\section{Preliminaries}
\label{sec: preliminaries}

We now give formal definitions. Let $G=(V,E)$ be an undirected graph.
Throughout the paper, we use $n = |V|$ to denote the number
of vertices in $G$ and $m= |E|$ to be the number of edges.
We will typically use letters $u,v,x,y,...$, possibly with
indices, to denote vertices, and $a,b,e,f,...$ to denote
edges. If an edge $e$ has endpoints $x,y$, we write $ e =
\braced{x,y}$. We allow multiple edges and loops in $G$, so
the endpoints do not uniquely define an edge:
if $e = \braced{x,y}$ and $f = \braced{x,y}$, it is not 
necessarily true that $e = f$.

A \emph{circulation} on $G$ is a function $\psi$ that
assigns a flow value and a direction to any edge in $E$. (We
use the terms ``circulation" and ``flow" interchangeably,
slightly abusing the terminology.)  Denoting by $\psi(u,v)$
the flow on $e= \braced{u,v}$ from $u$ to $v$, we require that
$\psi$ satisfies the following two conditions (i) $\psi$ is
anti-symmetric, that is $\psi(u,v) = -\psi(v,u)$ for each
edge $\braced{u,v}$, and (ii) $\psi$ satisfies the flow
conservation property, that is $\sum_{\braced{u,v}\in E}
\psi(u,v) = 0$ for each vertex $v$.

A \emph{bridge} in $G$ is an edge whose removal increases
the number of connected components of $G$.  Let
$\bridges(G)$ be the set of bridges in $G$.  

Suppose that some circulation $\psi$ is given for all edges
in some set $M \subseteq E$, and not for other edges.
We have the following observation:

\begin{observation}
For $\braced{u,v}\in E-M$, $\psi(u,v)$ is uniquely
determined from the flow preservation property
if and only if $\braced{u,v} \in\bridges(G-M)$.
\end{observation}

We can now define the \emph{gain} of $M$ to be $\gain(G,M) =
|M \cup \bridges(G-M)|$, that is, the total number of edges
for which the flow can be determined if we place monitors on
the edges in $M$. We will refer to the edges in $M$ as
\emph{monitor} edges, while the bridge edges in
$\bridges(G-M)$ will be called \emph{extra} edges. If $G$ is
understood from context, we will write simply $\gain(M)$
instead of $\gain(G,M)$.

\smallskip

The \emph{Flow Edge-Monitor Problem} ({\FlowMntrs}) can now be defined
formally as follows: given a graph $G = (V,E)$ and an
integer $k >0$, find a set $M\subseteq E$ with $|M| \le k$ that
maximizes $\gain(G,M)$.

%%%%%%%%%%%%%%%%%%%%%%%%%%

\paragraph{The weighted case.} 
We consider the extension of {\FlowMntrs} to weighted graphs, where each
edge $e$ has a non-negative weight $w(e)$ assigned to it,
and the task is to maximize the \emph{weighted gain}. More
precisely, if $M$ are the monitor edges, then the formula
for the (weighted) gain is $\gain(M) =
\sum_{e\in M\cup B} w(e)$, for $B = \bridges(G-M)$.
We will denote this problem by {\WFlowMntrs}.

Throughout the paper, we denote by
$M^\ast$ some arbitrary, but fixed, optimal monitor
edge set. Let $B^\ast = \bridges(G-M^\ast)$ be the
set of extra edges corresponding to $M^\ast$. 
Then the optimal gain is $\optgain(G,k) = w(M^\ast\cup B^\ast)$.

%%%%%%%%%%%%%%%%

\paragraph{Simplifying assumptions.} 
We make some assumptions about
the input graph $G$ that will simplify the proofs. First,
if $k\ge m$, then we can simply take $M=E$ and this will be an
optimal solution to {\WFlowMntrs}. Therefore, without loss of
generality, throughout the paper we will assume that $m > k$.

The flow value on any bridge of $G$ must be $0$, so we can assume
that $G$ does not have any bridges. Further, if $G$ is not 
connected, we can do this: take any two vertices $u$, $v$ from
different connected components and contract them into one vertex.
This operation does not affect the solution to {\WFlowMntrs}.
By repeating it enough many times, we can transform $G$ into a
connected graph.
Summarizing, we conclude that, without loss of generality, we
can assume that $G$ is connected and does not have any
bridges. In other words, $G$ is $2$-edge-connected. 
(Recall that, for an integer $c\ge 1$, a graph $H$ is called
$c$-edge-connected, if $H$ is connected and it remains
connected after removing any set of at most $c-1$ edges from $H$.)

\smallskip

Next, we claim that can in fact restrict our
attention to $3$-edge-connected graphs. To justify it, we
show that any weighted 2-edge-connected graph $G = (V,E)$ can be 
converted in linear time into a 3-edge-connected weighted graph $G'=
(V',E')$ such that:

\begin{description}
	\item{(i)} $\optgain(G,k) = \optgain(G',k)$, and
	\item{(ii)} If $M'\subseteq E'$ is a set of $k$ monitor
    edges in $G'$, then in linear time one can find a set
    $M\subseteq E$ of $k$ monitor edges in $G$ with
    $\gain(G,M) = \gain(G',M')$.
\end{description}

We now show the construction of $G'$.
A \emph{$2$-cut} is a pair of edges $\braced{e,e'}$ whose removal
disconnects $G$. Write $e \simeq e'$ if $\braced{e,e'}$ is
a $2$-cut. It is known, and quite easy to show, that
relation ``$\simeq$" is an equivalence relation on $E$.
The equivalence classes of $\simeq$ are called \emph{edge groups}.

\begin{figure}[ht]
\hfill
\noindent
    \begin{minipage}[c]{0.7\linewidth}
	\includegraphics[width=\linewidth]{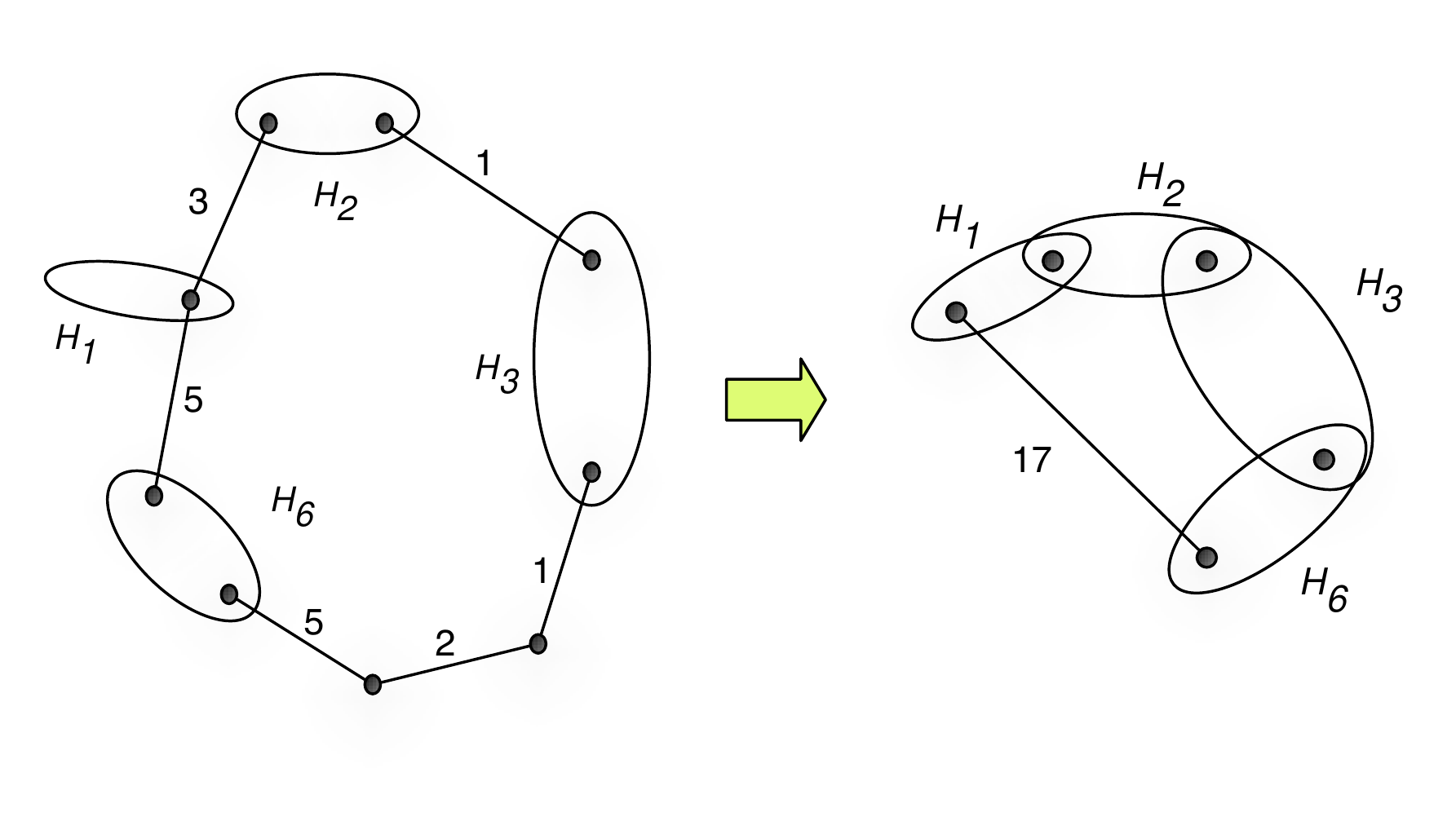}
    \end{minipage}
\hskip 0.15in
    \begin{minipage}[c]{0.2\linewidth}
    	\caption{Contracting edge groups.}
		\label{fig: contracting edge groups}
    \end{minipage}
\hfill{\ }
\end{figure}

Suppose that $G$ has an edge group $F$ with $|F|=q$, for $q\ge 2$,
and let $H_1,...,H_q$ be the connected components of $G-F$. 
Then $F = \braced{e_1,...,e_q}$, where, for each $i$,
$e_i =\braced{u_i,v_i}$, $u_i\in H_i$ and $v_i\in H_{i+1}$
(for $i=q$ we assume $q+1\equiv 1$).  
For $i= 1,...,q-1$, contract edge $e_i$ so that
vertices $u_i$ and $v_i$ become one vertex, and then assign
to edge $e_q = \braced{u_q,v_q}$ weight $\sum_{i=1}^q
w(e_i)$. We will refer to $e_q$ as the \emph{deputy edge}
for $F$. Figure~\ref{fig: contracting edge groups}
illustrates the construction.

%%%%%%%%%%%%

%%%%%%%%%%%%

Let $G'=(V',E')$ be the resulting weighted graph.  By the
construction, $G'$ is $3$-edge-connected.  All edge groups
can be computed in linear time (see, \cite{Tsin3edge}, for
example), so the whole transformation can be done in linear
time as well.

\smallskip

It remains to show that $G'$ satisfies conditions (i) and (ii).
If $M$ is any monitor set,
and if $M$ has two or more monitors in the same edge group, we can
remove one of these monitors without decreasing the gain of $M$.
Further, for any monitor edge $e$ of $M$, we can replace $e$
by the deputy edge of the edge group containing $e$, without
changing the gain. 
This implies that, without loss of generality, we can assume
that the optimal monitor set $M^\ast$ in $G$ consists only of
deputy edges. These edges remain in $G'$ and the gain of $M^\ast$
in $G'$ will be exactly the same as its gain in $G$.
This shows the ``$\ge$" inequality in (i). The ``$\le$"
inequality follows from the fact that any monitor set in $G'$
consists only of deputy edges from $G$.
The same argument implies (ii) as well.

\smallskip

Summarizing, we have shown in this section that, without loss of generality,
we can assume that the input graph $G$ has $m > k$ edges
and is $3$-edge-connected.

%%%%%%%%%%%%%%%%%%%%%%%%%%%%%%%%%%%%

\paragraph{The kernel graph.}
Consider an input graph $G=(V,E)$ and a monitor edge set $M$, and let
$B = \bridges(G-M)$.  The \emph{kernel graph associated with $G$ and $M$}
is defined as the weighted graph
$G_M=(V_M,E_M)$, where $V_M$ is the set of connected
components of $G - M - B$, and $E_M$ is determined as
follows: For any edge $e\in M \cup B$, where 
$e = \braced{u,v}$, let $u'$ and $v'$ be
the connected components of $G-M-B$ that contain,
respectively, $u$ and $v$. Then we add edge $\braced{u',v'}$ to
$E_M$. The weights are preserved, that is
$w(\braced{u',v'}) = w(\braced{u,v})$.
We will say that this edge $\braced{u',v'}$
\emph{represents} $\braced{u,v}$ or \emph{corresponds to}
$\braced{u,v}$. In fact, we will often
identify $\braced{u,v}$ with $\braced{u',v'}$, treating them
as the same object. We point out that in general $G_M$ is a multigraph, 
as it may have multiple edges and loops (even when $G$ does not).
However, since $G$ is $3$-edge-connected, it is easy to see that so is $G_M$.

Figure~\ref{fig: kernel graph example} shows the kernel graph corresponding to 
the graph and the monitor set in the example from 
Figure~\ref{fig: graph monitors example} (all edge weights are $1$):

\begin{figure}[ht]
\hfill
    \begin{minipage}[c]{0.55\linewidth}
	\includegraphics[width=\linewidth]{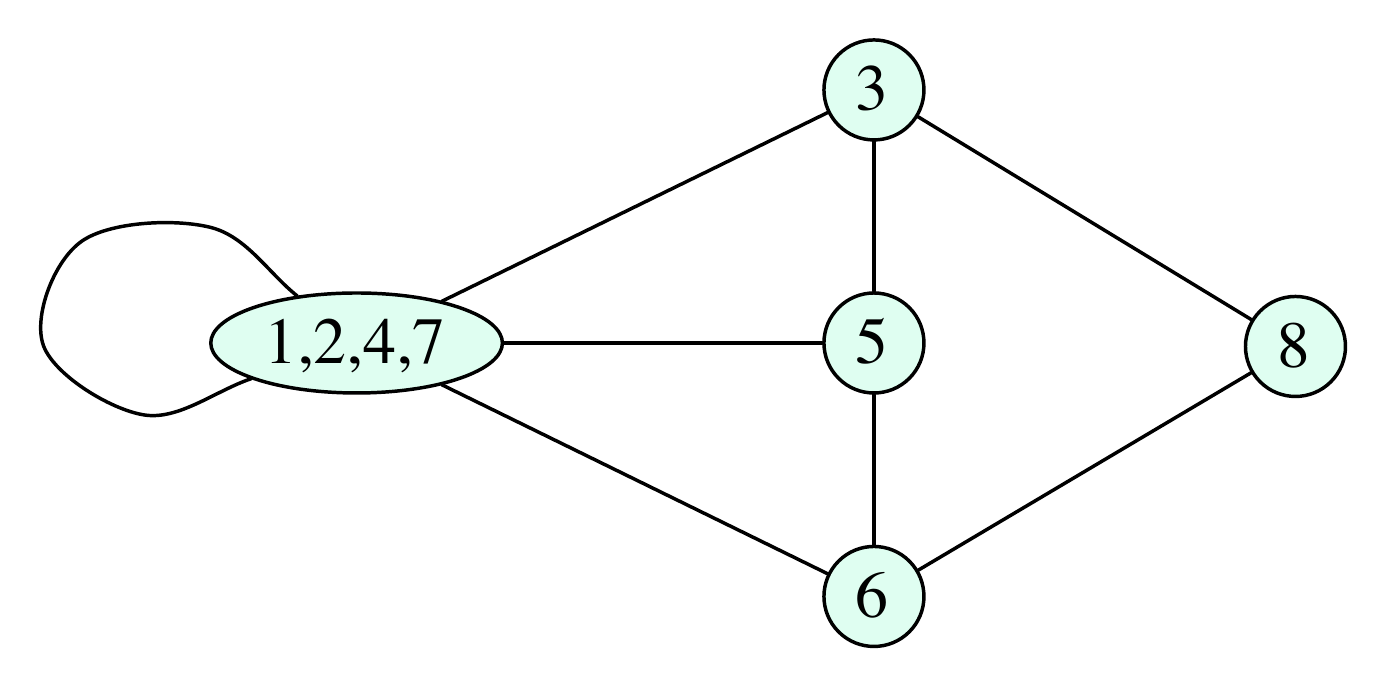}
    \end{minipage}
\hskip 0.15in
    \begin{minipage}[c]{0.35\linewidth}
    	\caption{The kernel graph for the example
 				in Figure~\ref{fig: graph monitors example}.
				The loop in vertex $\braced{1,2,4,7}$
				represents edge $\braced{2,1}$.}
		\label{fig: kernel graph example}
    \end{minipage}
\hfill{\ }
\end{figure}

Note that we have $|E_{M}| \le k + |V_{M}| - 1$.
This can be derived directly from the definitions: The
edges in $G_M$ that represent extra edges are the
bridges in $G_M$ and therefore they form a
forest in $G_M$. This (and the fact that $G_M$ is connected) implies that 
the number of extra edges is at most $|V_M|-1$, and the inequality follows.

\smallskip

In the paper, we will use the concept of kernel graphs only
with respect to some optimal monitor set.
Let $M^\ast$ be some arbitrary, but fixed, optimal monitor
edge set. To simplify notation, we will 
write $G^\ast = (V^\ast,E^\ast)$ for the kernel graph
associated with $M^\ast$, that is
$G^\ast = G_{M^\ast}$, $V^\ast = V_{M^\ast}$ and $E^\ast = E_{M^\ast}$.
In this notation, we have $\optgain(G,k) = w(E^\ast)$. 
In the analysis of our algorithms, we will be
comparing the weights of edges collected
by the algorithm against the edges in the kernel graph $G^\ast$.

%%%%%%%%%%%%%%%%%%%%%%%%%%%%%%%%%%%%%%%%%%%%%%%%%%%%%%%%%%%%%%%%%%%%%%
%%%%%%%%%%%%%%%%%%%%%%%%%%%%%%%%%%%%%%%%%%%%%%%%%%%%%%%%%%%%%%%%%%%%%%
%%%%%%%%%%%%%%%%%%%%%%%%%%%%%%%%%%%%%%%%%%%%%%%%%%%%%%%%%%%%%%%%%%%%%%

\section {Proof of NP-hardness of {\FlowMntrs}}

We show that the {\FlowMntrs} is {\NP}-hard
(even in the unweighted case), via a reduction from the
{\Clique} problem. We start with a simple lemma.

%%%%%%%%%%%

\begin{lemma} \label{lem: max_choose2} 
Let $a_1, a_2, \ldots, a_s$ be $s$ positive integers such
that $\sum_{i=1}^s a_i = n$, for a fixed integer $n$.
Then $\sum_{i=1}^s \binom{a_i}{2}$ is maximized if and
only if $a_j =
n-s+1$ for some $j$ and $a_i = 1$ for all $i\neq j$.
\end{lemma}

Above, we assume that $\binom{1}{2}= 1(1-0)/2 = 0$.

\begin{proof}
By routine algebra, one can verify that for $2\le a\le b$ we have
\begin{eqnarray}
	\binom{a}{2} + \binom{b}{2} &<& \binom{a-1}{2} + \binom{b+1}{2}.
	\label{eqn: np-hard, binom}
\end{eqnarray}
Without loss of generality, assume $a_1 = \max_i a_i$. If
$a_i > 1$, for any $i > 1$, by the inequality above, we can
change $a_1 \assign a_1+1$ and $a_i \assign a_i-1$,
increasing the value of $\sum_{i=1}^s \binom{a_i}{2}$. By
repeating this argument, we obtain that an optimum is
achieved for $a_1 = n-s+1$ and $a_2 = a_3 = ... = a_s = 1$.
That all optima have this form (up to a permutation of the
$a_i$'s) follows from the fact
that inequality (\ref{eqn: np-hard, binom}) is strict.
\end{proof}

\medskip

%%%%%%%%%%%%%%

\begin{theorem} \label{theorem:np_hard} 
{\FlowMntrs} is {\NP}-hard.
\end{theorem}

\begin{proof}
 In the {\Clique} problem, given an undirected graph
  $G=(V,E)$ and an integer $q>0$, we wish to
  determine if $G$ has a clique of size at least $q$. 
{\Clique} is well-known to be {\NP}-complete (see \cite{garey79}).
We show how to reduce {\Clique}, in polynomial-time, to {\DecFlowMntrs},
 the decision version of {\FlowMntrs}, 
defined as follows: Given a graph $G
  = (V,E)$ and two integers, $k,l > 0$, is there a set $M$
  of $k$ edges in $G$ for which $|\bridges(G-M)|\ge l$?

  The reduction is simple. Suppose we have an instance
  $G=(V,E),q$ of {\Clique}. Without
  loss of generality, we can assume that $G$ is connected
  and $q\ge 3$. Let $n = |V|$ and $m = |E|$.  We map this
  instance into an instance $G,k,l$ of {\DecFlowMntrs},
  where $k = m-\binom{q}{2} - l$ and $l=n-q$. This clearly takes
  polynomial time. Thus, to complete the proof, it is
  sufficient to prove the following claim:

\begin{description}
	\item{$(\ast)$}
 $G$ has a clique of size $q$ iff $G$ has
a set $M$ of $k$ edges for which $|\bridges(G-M)| \ge l$.
\end{description}

We now prove $(\ast)$. The main idea is that, by the choice of parameters
$k$ and $l$, the monitors and extra edges in the solution
of the instance of {\DecFlowMntrs} must be exactly the
edges outside the size-$q$ clique of $G$.

\smallskip

$(\Rightarrow)$ 
Suppose that $G$ has a clique $C$ of size $q$. Let $G'$ be
the graph obtained by contracting $C$ into a single vertex and
let $T$ be a spanning tree of $G'$. We then take $M$ to be the
set of edges of $G'$ outside $T$. Thus the edges in $T$
will be the bridges of $G-M$.
Since $G'$ has $n-q+1$ vertices, $T$ has $l = n-q$ edges, and
$M$ has $m-\binom{q}{2} - l = k$ edges.

\smallskip

$(\Leftarrow)$ Suppose there is a set $M$ of $k$ monitor
edges that yields a set $B$ of $l'$ extra edges, where $l\le l'\le n-1$.
We show that $G$ has a clique of size $q$.

Let $s$ be the number of connected components of $G-M-B$,
and denote by $a_1,a_2,...,a_s$ the cardinalities of these
components (numbers of vertices). Since $|B|=l'$, we have
$s\ge l'+1$.  Also, $\sum_{i=1}^s a_i = n$ and $\sum_{i=1}^s
\binom{a_i}{2} + k + l' \ge m$. Therefore, using
Lemma~\ref{lem: max_choose2}, and the choice of $k$ and $l$,
we have
\begin{eqnarray}
	\binom{n-l'}{2} + l' &\ge&  \binom{n-s+1}{2} + l'
			\label{eqn: np-hard derivation 1} \\
			&\ge& 
  			\sum_{i=1}^s \binom{a_i}{2} + l'
			\label{eqn: np-hard derivation 2}\\
			&\ge& m- k 
			\label{eqn: np-hard derivation 3}\\
			&=& \binom{n-l}{2} + l. 
						\nonumber
\end{eqnarray}
By routine calculus, the function $f(x) = \half (n-x)(n-x-1) + x$
is decreasing in interval $[0,n-1]$, and therefore the above
derivation implies that $l'\le l$, so we can conclude that $l' = l$.
This, in turn, implies that all inequalities in this derivation
are in fact equalities. Since
(\ref{eqn: np-hard derivation 1}) is an equality, we have 
$s-1 = l' = l = n-q$.
Then, since (\ref{eqn: np-hard derivation 2}) is an equality,
Lemma~\ref{lem: max_choose2} implies that $a_j = q$ for some $j$ and
$a_i = 1$ for all $i\neq j$.
Finally, (\ref{eqn: np-hard derivation 3}) can be an equality only
if all the connected components are cliques. In particular,
we obtain that the $j$th component is a clique of size $q$.
\end{proof}

%%%%%%%%%%%%%%%%%%%%%%%%%%%%%%%%%%%%%%%%%%%%%%%%%%%%%%%%%%%%%%%%%%%%%%
%%%%%%%%%%%%%%%%%%%%%%%%%%%%%%%%%%%%%%%%%%%%%%%%%%%%%%%%%%%%%%%%%%%%%%
%%%%%%%%%%%%%%%%%%%%%%%%%%%%%%%%%%%%%%%%%%%%%%%%%%%%%%%%%%%%%%%%%%%%%%

\section{Algorithm~{$\sigma$-\Greedy}}
\label{sec: sigma-Greedy}

Fix some integer constant $\sigma\ge 1$. Let $G = (V,E)$ be
the input graph with $n = |V|$, $m = |E|$, and with
weights on edges. As justified in 
Section~\ref{sec: preliminaries}, we will assume that $m > k$
and that $G$ is $3$-edge-connected.

Algorithm~{$\sigma$-\Greedy} that we study in this section
works in $\ceiling{k/\sigma}$ steps and returns a set of $k$
monitor edges. In each step, it assigns $\sigma$ monitors to
a set $P$ of $\sigma$ edges that maximizes the gain in
this step, that is, the total weight of the monitor edges in $P$
and the bridges in $G-P$. These edges are then removed from $G$,
and the process is repeated. A more rigorous description is
given in Figure~\ref{fig: sigma-greedy}, which also deals
with special cases when the number of monitors or edges left
in the graph is less than $\sigma$.

\begin{figure}[ht]
\begin{center}
\begin{minipage}{4in}
{
\input{pseudocode.tex}
\begin{program}
Algorithm~{$\sigma$-\Greedy}
   |$G_0 = (V,E_0) \assign G = (V,E)$|
   |$M_0 \assign \emptyset$|
   |$X_0 \assign \emptyset$|
   for |$t \assign 1,2,...,\ceiling{k/\sigma}$|
       if |$E_{t-1} = \emptyset$|
           then |return $M = M_{t-1}$ and halt|
       |$\sigma' \assign \sigma$|
       if |$t = \floor{k/\sigma}+1$|
           then |$\sigma' = {k \bmod \sigma}$|
       if |$|E_{t-1}| \le \sigma'$|
           then |$P \assign E_{t-1}$|
           else
               |find $P\subseteq E_{t-1}$ with $|P|=\sigma'$|
                  |that maximizes $w(P\cup \bridges(G_{t-1}-P))$|
               |$Y_t \assign P\cup \bridges(G_{t-1}-P)$|
               |$X_t \assign X_{t-1} \cup Y_t$|
               |$E_t \assign E_{t-1}-Y_t$|
               |$G_t \assign (V,E_t)$|
       |$M_t \assign M_{t-1} \cup P$| %% Comment, this line is done in every iteration

   |return $M = M_{\ceiling{k/\sigma}}$|
\end{program}
}
\end{minipage}
\end{center}
\caption{Pseudo-code for Algorithm~{$\sigma$-\Greedy}. $Y_t$
represents the edges collected by the algorithm in step $t$,
with $P\subseteq Y_t$ being the set of monitor edges and
$Y_t-P$ the set of extra edges. $M_t$ represents all monitor edges
collected up to step $t$ and $X_t$ represents all edges 
collected up to step $t$.}
\label{fig: sigma-greedy}
\end{figure}

Note that each step of the algorithm runs in time
$O(m^\sigma(n+m))$, by trying all possible combinations
of $\sigma$ edges in the remaining graph $G_{t-1}$ to find
$P$.  Hence, for each fixed $\sigma$,
Algorithm~{$\sigma$-\Greedy} runs in polynomial time.

%%%%%%%%%%%%%%%%%%%%%%%%%%%%%%%%%%%%%%%%%%%%%%%%%%%%%%%%%%%%%%%%%%%%%%
%%%%%%%%%%%%%%%%%%%%%%%%%%%%%%%%%%%%%%%%%%%%%%%%%%%%%%%%%%%%%%%%%%%%%%

\subsection{Analysis of {\OneGreedy}}
\label{subs: onegreedy}

In this section we consider the case $\sigma=1$.
Algorithm~{\OneGreedy} at each step
chooses an edge whose removal maximizes the gain
and places a monitor on this edge. This edge and its
corresponding bridges are removed from the graph.
This process is repeated $k$ times. We show that this 
algorithm has approximation ratio $3$.

%%%%%%%%%%%%

\paragraph{Analysis.}
Fix the value of $k$, and some optimal solution $M^\ast$ of $k$
monitor edges, and let $\resG = (\resV,\resE)$ be the corresponding 
kernel graph. To avoid
clutter, we will identify each edge in $\resE$
with its corresponding edge in $E$, thus thinking of $\resE$
as a subset of $E$.  For example, when we say that the
algorithm collected some $e\in \resE$, we mean that it
collected the edge in $E$ represented by $e$.

Recall that $\optgain(G,k) = w(\resE)$, where
$w(\resE)$ is the sum of weights of the edges in
$\resE$. Thus we need to show that {\OneGreedy}'s gain is
at least $\onethird w(\resE)$.

Let $e_i$, $i= 1,2,...,k$, be the $k$ heaviest edges in $\resE$,
ordered by weight, that is $w(e_1) \ge w(e_2) \ge ... \ge w(e_k)$.
First, we claim that for each $ t= 0,1,...,k$, we have
\begin{eqnarray}
 	w(X_t) &\ge& {\textstyle \sum_{i=1}^t w(e_i)},
		\label{eqn: induction 1-greedy}
\end{eqnarray}
where $X_t$ denotes the set of edges collected by the algorithm
in the first $t$ steps.

The proof of (\ref{eqn: induction 1-greedy}) is by a 
straightforward induction. It is vacuously true for $t=0$. Suppose
(\ref{eqn: induction 1-greedy}) holds for $t' = t-1$; we will
then show that it also holds for $t$. If $X_t$ contains
all edges $e_1,...,e_t$, then (\ref{eqn: induction 1-greedy}) holds.
Otherwise, choose any $j$, $1\le j\le t$, for which $e_j\notin X_t$.
By induction, $w(X_{t-1})\ge \sum_{i=1}^{t-1} w(e_i)$, and $e_j$ is
available to {\OneGreedy} in step $t$, so its gain in step $t$
is at least $w(e_j)$. Therefore 
$w(X_{t})\ge w(X_{t-1}) + w(e_j) 
  			\ge w(X_{t-1}) + w(e_t) \ge	\sum_{i=1}^{t} w(e_i)$,
completing the proof of (\ref{eqn: induction 1-greedy}).

\smallskip

Since $G$ is $3$-edge-connected, each vertex in
$\resG$ has degree at least $3$, so $|\resE| \ge \frac{3}{2}
|\resV|$, which implies that $k \ge |\resE| - |\resV| + 1 > \onethird
|\resE|$. Thus, from (\ref{eqn: induction 1-greedy}),
we obtain that the gain of  {\OneGreedy} is
$w(X_k) \ge \sum_{i=1}^k w(e_i) \ge \onethird w(\resE)$.
Summarizing, we obtain:

%%%%%%%%%%%%%%

\begin{theorem}\label{eqn: onegreedy}
Algorithm~{\OneGreedy} is a polynomial-time
$3$-approximation algorithm for the Weighted
Flow Edge-Monitor Problem, {\WFlowMntrs}.
\end{theorem}

With a somewhat more careful analysis, one can show that the
approximation ratio of {\OneGreedy} is actually $3(1-1/k)$,
which matches our lower bound example below. 

%%%%%%

\begin{figure}
\hfill
\noindent
    \begin{minipage}[c]{0.6\linewidth}
	\includegraphics[width=\linewidth]{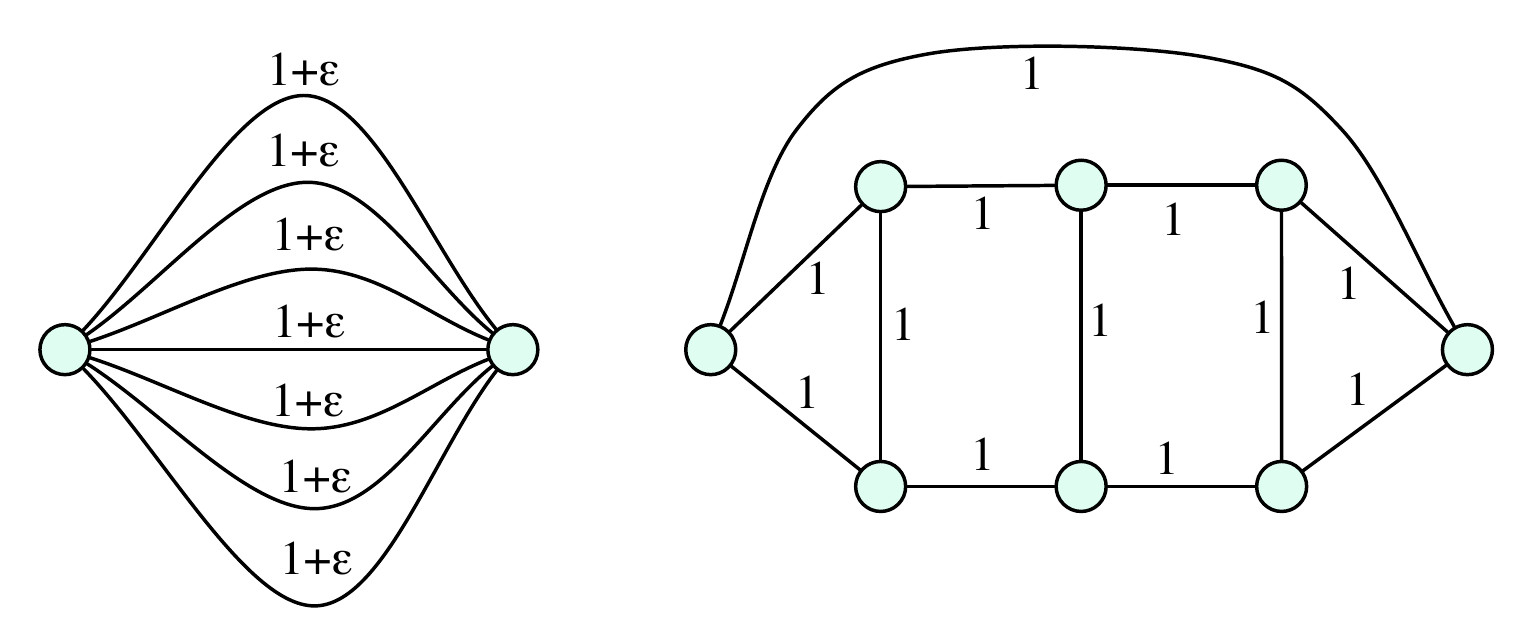}
    \end{minipage}
%
%\hskip 0.15in
%
    \begin{minipage}[c]{0.35\linewidth}
    	\caption{Lower bound example for {\OneGreedy}, with $k=5$.}
		\label{fig: lower bound for 1-greedy}
    \end{minipage}
\hfill{\ }
\end{figure}

\paragraph{A tight-bound example.}
We now present an example showing that our analysis of
{\OneGreedy} is tight. Graph $G$ consists
of one connected component with $2k-2$ vertices, in which
each vertex has degree $3$ and each edge has weight $1$, and
the other connected component that has only two vertices
connected by $k+2$ edges each of weight $1+\epsilon$.
Fig.~\ref{fig: lower bound for 1-greedy} shows the
construction for $k=5$.

%%%%%%%%%%%%

%%%%%%%%%%%%

{\OneGreedy} will be collecting edges from the $2$-vertex
component on the left, ending up with $k$ edges and total
gain $(1+\epsilon) k$. The optimum solution is to put $k$
monitors in the cubic component on the right, thus gaining
all $3k-3$ edges from this component. For $\epsilon\to 0$,
the approximation ratio tends to $3(1-1/k)$.

%%%%%%%%%%%%%%%%%%%%%%%%%%%%%%%%%%%%%%%%%%%%%%%%%%%%%%%%%%%%%%%%%%%%%%
%%%%%%%%%%%%%%%%%%%%%%%%%%%%%%%%%%%%%%%%%%%%%%%%%%%%%%%%%%%%%%%%%%%%%%

\subsection{Analysis of {\TwoGreedy}}
\label{subs: twogreedy}

We now consider $\sigma = 2$.  At each step,
Algorithm~{\TwoGreedy} selects two edges that maximize the
gain. Ties are broken arbitrarily. We place monitors on
these two edges, and then remove them from $G$, as well as
the resulting bridges.  The process continues for
$\floor{\half k}$ steps.  Exceptional situations, when $k$
is odd, or we run out of edges, etc., are handled as in
Figure~\ref{fig: sigma-greedy}.  In this section we show
that {\TwoGreedy} is a $2$-approximation algorithm.

%%%%%%%%%%%%%%%%%%%%%%%%%%%%%%%%%%%%%

\paragraph{Idea of the analysis.}
Let $G = (V,E)$ be the input graph with weights on edges. As
before, we will assume that $m>k$ and that $G$ is $3$-edge-connected.
We can further assume that {\TwoGreedy} never runs out of edges,
for otherwise it computes an optimal solution.
We fix some optimal solution $M^{\ast}$, and
let $\resG = (\resV,\resE)$ be the corresponding kernel
graph with $\nu = |\resV|$ vertices and $\mu = |\resE|$ edges.
Recall that $\optgain(G,k) = w(\resE)$; thus we need
to show that {\TwoGreedy}'s gain is at least $\half w(\resE)$.

Our proof for {\OneGreedy} was based on the observation that
$k\ge \onethird |\resE|$; thus we only
needed to show that {\OneGreedy}'s gain is at least the
total weight of the $k$ heaviest edges in $\resE$. This is
not sufficient for {\TwoGreedy}. To see this, consider a
simple situation where $G^\ast$ is unweighted with all 
vertices of degree $3$, the extra edges form a tree
in $\resG$, and $k$ is even. Then $\mu = \threehalfs \nu$
and $\mu = \nu+k-1$, so $\mu\approx 3k$.  Therefore we need
to show that in this case {\TwoGreedy} collects at least
$\threehalfs k$ edges. For the unweighted graph, this is not hard
to show: pick a set $J$ of $\half k$ independent vertices in
$\resG$. For each vertex $v$ in $J$, at each step of
{\TwoGreedy}, either the three edges of $v$ have already
been collected, or they can be collected in this step by
placing monitors on two of them. (It is also possible that
one edge out of $v$ has already been collected, but in this
case the remaining two can be collected in this step as
well.) Therefore the gain of {\TwoGreedy} in $\half k$ steps
will be at least the number of edges out of $J$, that is
$\threehalfs k$.

If $G$ is weighted, the argument above is not sufficient,
since the edges incident to vertices in $J$ may have small
weights. Further, it \emph{may} happen that {\TwoGreedy}
will collect only $k$ edges in total, if they are heavy
enough, and in this case we need to argue that their total
weight is at least $\half w(\resE)$, even though $k$ may be
only about $\onethird |\resE|$.

The proof consists of two parts. First, we introduce the
concept of a \emph{bundle}. Intuitively, a bundle is a set
of edges that can be collected with at most two monitors,
although our definition is more restrictive and will be
given shortly. We show that $\resG$ contains a set $T$ of at
most $\floor{\half k}$ disjoint bundles with total weight at
least $\half w(\resE)$. In the second part of the proof we
show that the gain of {\TwoGreedy} is at least the total
weight of $T$.

%%%%%%%%%%%%

\paragraph{Analysis.}
For any vertex in $\resG$ of degree $3$, the set of three edges
incident on this vertex is called a \emph{tripod}.
A set $\beta$ of edges is called a \emph{bundle} if $\beta$ is either
a tripod or it consists of at most two edges. Clearly, all edges
of a bundle can be collected with at most $2$ monitors.
If $T$ is a set of bundles, by $E_T$ we denote the set of
edges in $T$, that is $E_T = \bigcup_{\beta\in T} \beta$.
(We will extend the definition of bundles later in the proof of
Lemma~\ref{lem: 2greedy, alg >= target}.)

%%%%%%%%%%%%%%

\begin{lemma} \label{lem: 2greedy, construct target} 
  For $k \ge 2$, there exists a set $T$ of at most
  $\floor{k/2}$ disjoint bundles such that $w(E_T)\ge
  \half w(\resE)$.
\end{lemma}

\begin{proof}
  First, we construct a collection $Z$ of bundles that
  contains all tripods in $\resG$ and such that each edge
  appears in exactly two bundles in $Z$. To this end, we
  create two copies of each edge.  If $e = \braced{x,y}$ is
  an edge, then by $e^x$ and $e^y$ we will denote these two
  copies of $e$. Let $F$ be the set of all these edge copies.

  For each vertex $v$ of $\resG$ of degree $3$, if $e$, $f$,
  $g$ are the edges incident to $v$, add the tripod $\beta =
  \braced{e^v,f^v,g^v}$ to $Z$ and remove $e^v$, $f^v$ and
  $g^v$ from $F$. Let $F'$ be the set of remaining edges in $F$.

  We now want to partition  $F'$ into
  pairs, with each pair becoming a bundle. Assume first that
  $|F'|$ is even; the argument for the general case is
  essentially the same but slightly more cumbersome because
  of the parity issue, and it will be explained later.

  Group the edges in $F'$ arbitrarily into pairs. As long as any of these
  pairs has two copies of the same edge, say
  $\braced{e^x,e^y}$, do this: take any other pair
  $\braced{f^u,g^v}$ (where possibly $f=g$) and replace
  these two pairs by pairs $\braced{e^x,f^u}$ and
  $\braced{e^y,g^v}$.  Eventually each pair will contain
  different edges. All these pairs
  become bundles that are now added to $Z$. 
 This completes the construction of $Z$. 

 To construct $T$, we now greedily
  extract from $Z$ non-overlapping bundles with maximum
  weight.  More specifically, we start with $T = \emptyset$,
  and repeat the following step as long as $|Z|\ge 4$:
  choose a bundle $\beta$ in $Z$ with maximum $w(\beta)$,
  add it to $T$, and then remove from $Z$ exactly four
  bundles: $\beta$, all other bundles in $Z$ that intersect
  $\beta$, and possible a few more additional bundles so
  that the total is four. This is possible, because, by the
  definition of $Z$, each bundle in $Z$ intersects at most
  three other bundles in $Z$. If, at the end, $|Z|\neq\emptyset$, then we add to
  $T$ the bundle $\beta$ in $Z$ with maximum $w(\beta)$ and
  remove all remaining bundles (at most three) from $Z$.

  We now have our set $T$ of bundles, and it remains to show
  that it has the desired properties. Obviously, all bundles
  in $T$ are disjoint.  In each step of the construction of
  $T$, when we add a bundle $\beta$ to $T$, we reduce
  $w(E_Z)$ by at most $4w(\beta)$, so $w(E_T) \ge \onefourth
  w(E_Z) = \half w(\resE)$, as needed.

  It remains to show that $|T|\le \floor{\half k}$.  Let
  $\nu_d$ be the number of vertices of degree $d$ in
  $\resG$.  All vertices in $\resG$ have degree at least
  $3$, thus
\begin{eqnarray*}
2\mu \;=\; {\textstyle \sum_{d\geq 3} d\cdot \nu_d }
     \;\geq\; 3\nu_3 + 4(\nu-\nu_3) 
    \;=\; 4\nu-\nu_3. 
\end{eqnarray*}
We also have $\mu\le \nu+k-1$, which, together with the
inequality above yields $2\mu - \nu_3 \le 4k-4$. 
In $Z$, we have exactly $\nu_3$ tripods and $\half(2\mu -
3\nu_3) = \mu - \threehalfs \nu_3$ pairs, so
we obtain $|Z| = \mu - \half\nu_3 \le
2k-2$.  Hence $|T| \leq \ceiling{(2k-2)/ 4} =
\ceiling{(k-1)/2} \le \floor{\half k}$, as needed.

Now we deal with the case when $|F'|$ is odd. We execute
the same process for pairing edges, but in this case
we will end up with one unpaired edge that will become a bundle
by itself. The proof that $w(E_T) \ge \half w(\resE)$ 
remains valid. We still need
to show that $|T| \leq \floor{\half k}$. In $Z$, we have
$\nu_3$ tripods, $\half(2\mu-3\nu_3-1)$ pairs, and one singleton;
hence $|Z| = \nu_3 + \half(2\mu-3\nu_3-1) + 1 = \half(2\mu-\nu_3+1)$.  
As before, we have $2\mu-\nu_3 \leq 4k-4$, but now
$2\mu - \nu_3 = (2\mu - 3\nu_3) + 2\nu_3 = |F'| + 2\nu_3$ is
odd, which implies that we in fact have 
$2\mu-\nu_3+1 \leq 4k-4$. This implies $|Z| \leq 2k-2$, and
the bound $|T|\le \floor{k/2}$ follows, as before.
\end{proof}

\medskip

%%%%%%%%%%%%%%

\begin{lemma} \label{lem: 2greedy, alg >= target} 
  Let $k\geq 2$, and let $T$ be the set of bundles
  constructed in Lemma~\ref{lem: 2greedy, construct
    target}. Then  $w(X_{\floor{k/2}}) \ge w(E_T)$;
thus {\TwoGreedy}'s gain is at least $w(E_T)$.
\end{lemma}

\begin{proof}
  Let $\ell = \floor{k/2}$.  To prove the lemma, we
show that there is a partition of $E_T$ into disjoint sets
  $B_1,B_2,...,B_\ell$ (some possibly empty) such that
  $w(B_t) \le w(Y_t)$, for $t = 1,2,...,\ell$, where $Y_t$
  is the set of edges collected by {\TwoGreedy} in step $t$.
  This is clearly sufficient to establish the theorem.

The idea of the proof is to proceed one step at a time,
for $t= 1,2,...,\ell$, starting with $T' = T$ and
at each step reducing the number of
bundles in $T'$. The invariant will be that at each step $t$,
all bundles in $T'$ will be available to {\TwoGreedy}
for collection. At step $t$, we eliminate from 
these bundles all edges collected by the algorithm, and rearrange
some bundles, if necessary, in such a way that the number
of bundles in $T'$ strictly decreases.

To implement the above idea we need to extend the definition
of bundles. If $\beta = \braced{e,f,g}$ is a tripod in $\resG$ and
{\TwoGreedy} collects $e$ at some step $t' < t$ while
$f,g$ remain uncollected until step $t$, then the pair
$\beta' = \braced{f,g}$ is called a \emph{bipod at step $t$}.
If $\beta\subseteq \resE$ is a set of edges not collected in the
first $t-1$ steps, then $\beta$ is called a 
\emph{bundle at step $t$} if either (i) $\beta$ is a tripod in
$\resG$, or (ii) $\beta = \beta_1\cup \beta_2$, where each
$\beta_i$ is either an edge or a bipod or $\emptyset$.
(We will omit phrase ``at time $t$" whenever $t$ is understood
from context.) Clearly, this definition extends the one given earlier
in this section. However, it is still true that {\TwoGreedy}
can collect all edges from a bundle in a single step (that is, with two
monitors).

Now we describe the construction of the sets $B_t$.
Initially, $T' = T$. Suppose that, for some $t\ge 1$, we have already
constructed $B_1,...,B_{t-1}$ and modified $T'$ accordingly. 
Consider now step $t$ of {\TwoGreedy}. We distinguish three cases.

\smallskip\noindent 
\mycase{1} $Y_t$ intersects at most one bundle in $T'$.
If $Y_t$ intersects one bundle, let $\beta$ be this bundle;
otherwise, let $\beta$ be any bundle. Remove
$\beta$ from $T$ and set $B_t =\beta$.

\smallskip\noindent
\mycase{2} $Y_t$ contains a bundle in $T'$. In this case we
simply set $B_t = Y_t\cap E_{T'}$ and we update $T'$ as follows: 
remove from $T'$ all bundles that are contained in $Y_t$,
and for each other bundle $\beta$ in $T'$ that intersects $Y_t$, 
remove from $\beta$ the edges in $Y_t$ (that is,
$\beta\assign\beta - Y_t$).

\smallskip\noindent 
\mycase{3} $Y_t$ intersects at least two bundles in $T'$ but
it does not contain any. We let $B_t = Y_t\cap E_{T'}$. To update $T'$,
we proceed in two steps. First,
choose any two bundles $\beta_1,\beta_2$ in $T'$ intersected by
$Y_t$, and replace them by $\beta_1\cup \beta_2 - Y_t$.
Then, for each other bundle $\beta$ in $T'$ intersected by $Y_t$, 
remove from $\beta$ all edges in $Y_t$. 

\smallskip
Note that Cases~2 and 3 are not mutually exclusive. If, at some
steps, both of these cases apply, any of them can be chosen
arbitrarily.

We claim that this procedure is correct, in the sense that at
each step $t$ all elements of $T'$ are bundles.
To this end, we make two observations. If $\beta = \braced{e,f,g}$ 
is a tripod at time $t$, then {\TwoGreedy} can either collect one edge
from $\beta$ or all three, but it cannot collect just two. If one
edge is collected, the remaining edges form a bipod. Also, if
$\braced{e,f}$ is a bipod at time $t$, then {\TwoGreedy} either collects
both edges $e,f$ or none. If we remove any edges from a bundle,
it obviously remains a bundle. Another type of update occurs in Case~3
where we replace $\beta_1$ and $\beta_2$ by $\beta_1\cup\beta_2- Y_t$.
Here, by the case condition, each $\beta_i-Y_t$ is either an edge or
a bipod, and thus $\beta_1\cup\beta_2- Y_t$ is a correct bundle.

Next, we estimate the weight of the sets $B_t$.
In Cases~2 and 3 we have $B_t \subseteq Y_t$, so clearly $w(Y_t)\geq
w(B_t)$. Due to the fact that it takes no more than two
monitors to collect all edges in a bundle and {\TwoGreedy} chooses $Y_t$,
$w(Y_t)$ is at least as large as the weight of any bundle in $T'$ at time $t$; 
hence $w(Y_t)\geq w(B_t)$ holds for Case~1 as well.

Finally, note that we have no more than $\ell$
bundles in $T'$ to start with and the total number of
bundles in $T'$ strictly decreases in each step. 
Therefore after $\ell$ steps we will have $T' = \emptyset$,
and thus $B_1,B_2,...,B_\ell$ is a partition of $E_T$,
completing the proof.
\end{proof}

In the case for $k=1$, {\TwoGreedy} actually gets an optimal
solution, and Lemma~\ref{lem: 2greedy, construct target} and
\ref{lem: 2greedy, alg >= target} combined imply that
for $k\geq 2$ the gain of {\TwoGreedy} is at least half of the
optimum. Thus, summarizing, we obtain our main result.

%%%%%%%%%%

\begin{theorem}
  Algorithm~{\TwoGreedy} is a polynomial-time
  $2$-approximation algorithm for the Weighted Flow
  Edge-Monitor Problem, {\WFlowMntrs}.
\end{theorem}

%%%%%%

\paragraph{A tight-bound example.}
Our analysis of {\TwoGreedy} is tight. The example is
essentially the same as the one for {\OneGreedy}, illustrated
in Figure~\ref{fig: lower bound for 1-greedy}, except that
the edges in the 2-vertex component on the left side have
now weights $1.5+\epsilon$. {\TwoGreedy} will be collecting
edges from this 2-vertex component, so its total gain will
be $(1.5+\epsilon)k$, while the optimum gain is $3k-3$.
For $\epsilon\to 0$ and $k\to\infty$, the ratio tends to $2$.

%%%%%%%%%%%%%%%%%%%%%%%%%%%%%%%%%%%%%%%%%%%%%%%%%%%%%%%%%%%%%%%%%%%%%%
%%%%%%%%%%%%%%%%%%%%%%%%%%%%%%%%%%%%%%%%%%%%%%%%%%%%%%%%%%%%%%%%%%%%%%
%%%%%%%%%%%%%%%%%%%%%%%%%%%%%%%%%%%%%%%%%%%%%%%%%%%%%%%%%%%%%%%%%%%%%%

\section{Final Comments}
\label{sec: final comments}

The most intriguing open question is what is the
approximation ratio of $\sigma$-{\Greedy} in the limit 
for $\sigma\to\infty$. 
We can show that this limit is not lower than $1.5$,
and we conjecture that $1.5$ is indeed the correct answer.

A natural question to ask is whether our results can be
extended to directed graphs. It is not difficult to show
that this is indeed true; both the {\NP}-hardness proof and
$2$-approximation can be adapted to that case.

Another intriguing direction to pursue would be to study
the extension of our problem to arbitrary linear systems of
equations. In that context, we can put $k$ ``monitors" on
$k$ variables of the system to measure their values. The
objective is to maximize the number of variables whose
values can be uniquely deduced from the monitored variables.

%%%%%%%%%%%%%%%%%%%%%%%%%%%%%%%%%%%%%%%%%%%%%%%%%%%%%%%%%%%%%%%%%%%%%%
%%%%%%%%%%%%%%%%%%%%%%%%%%%%%%%%%%%%%%%%%%%%%%%%%%%%%%%%%%%%%%%%%%%%%%
%%%%%%%%%%%%%%%%%%%%%%%%%%%%%%%%%%%%%%%%%%%%%%%%%%%%%%%%%%%%%%%%%%%%%%


\begin{thebibliography}{4}

\bibitem{DJPSY-mcuts-92} Dahlhaus, E. and Johnson, D.S., and
  Papadimitriou, C.H. and Seymour, P.D. and Yannakakis, M.:
  The Complexity of Multiway Cuts (Extended Abstract). In:
  24th ACM Symposium on Theory of Computing, STOC,
  pp. 241-251. ACM, New York (1992)

\bibitem{garey79} Garey, M.R. and Johnson, D.S.: Computers
  and Intractability: A Guide to the Theory of
  NP-Completeness. W. H. Freeman (1979)

\bibitem{GH-kcut-88} Goldschmidt, O and Hochbaum, D.S:
  Polynomial Algorithm for the k-Cut Problem. In: 29th
  Annual IEEE Symposium on Foundations of Computer Science,
  FOCS, pp. 444-451. IEEE Computer Society, Washington, DC,
  USA (1988)

\bibitem{GJ-traffic-05} Gu, W. and Jia X.: On a Traffic
  Control Problem. In: 8th International Symposium on
  Parallel Architectures,Algorithms and Networks , I-SPAN,
  pp. 510--515, IEEE Computer Society, Washington, DC, USA
  (2005)

\bibitem{KBP-meters-03} Khuller, S.  Bhatia, R. and Pless,
  R.: On Local Search and Placement of Meters in
  Networks. SIAM J on Comput. 32, 470-487 (2003)

\bibitem{SV-cuts-91} Saran, H. and Vazirani, V.V.: Finding
  k-cuts within Twice the Optimal. In: 32nd Annual Symposium
  on Foundations of Computer Science, FOCS,
  pp. 743-751. IEEE Computer Society, Washington, DC, USA
  (1991)

\bibitem{Tsin3edge} Tsin, Y.H.: A Simple 3-Edge-Connected
  Component Algorithm. Theor. Comp. Sys. 40, 125--142 (2007)

\bibitem{vazirani} Vazirani, V.V.: Approximation
  Algorithms. Springer (2001)

\end{thebibliography}
\end{document}